\documentclass[acmsmall,screen,nonacm]{acmart}

\setcopyright{none}
\acmDOI{}
\acmISBN{}

\usepackage{multirow}
\usepackage{enumitem}
\usepackage[table]{xcolor}
\usepackage[most]{tcolorbox}
\usepackage{listings}
\usepackage{pifont}
\usepackage{adjustbox}

\definecolor{modelblue}{HTML}{E0EBFF}
\definecolor{lightgrey}{HTML}{E6E6E6}

\definecolor{TblBlue}{RGB}{77,88,234}
\definecolor{TblBlueFill}{RGB}{235,241,250}
\definecolor{TblRed}{RGB}{196,59,59}
\definecolor{TblRedFill}{RGB}{252,239,239}

\newcommand{\tblimppos}[1]{\textcolor{TblRed}{\textbf{#1}}}

\definecolor{synKw}{HTML}{2563EB}
\definecolor{synStr}{HTML}{B45309}
\definecolor{synSelf}{HTML}{7C3AED}
\definecolor{synFn}{HTML}{0369A1}
\definecolor{synText}{HTML}{1F2328}
\definecolor{synCom}{HTML}{8B949E}

\lstdefinestyle{casecode}{
  language=Python,
  basicstyle=\ttfamily\scriptsize\color{synText},
  keywords={def,if,else,elif,for,in,is,not,and,or,return,import,None,True,False,while,continue},
  keywordstyle=\color{synKw},
  keywordstyle={[2]\color{synText}},
  keywordstyle={[3]\color{synText}},
  stringstyle=\color{synStr},
  commentstyle=\itshape\color{synCom},
  emph={self},
  emphstyle=\color{synSelf},
  emph={[2]next,test,from_string,__init__,_convert_string},
  emphstyle={[2]\color{synFn}},
  numbers=left,
  numberstyle=\tiny\color{gray},
  numbersep=4pt,
  xleftmargin=8pt,
  columns=fullflexible,
  keepspaces=true,
  showstringspaces=false,
  breaklines=false,
  aboveskip=0pt,
  belowskip=0pt,
}

\definecolor{passFill}{HTML}{EEF3EE}
\definecolor{passLine}{HTML}{A9C2AD}
\definecolor{passText}{HTML}{2F5D3A}

\definecolor{failFill}{HTML}{F5EDEB}
\definecolor{failLine}{HTML}{CFA9A3}
\definecolor{failText}{HTML}{8A2E26}

\newtcbox{\passbox}{
  on line,
  colback=passFill,
  colframe=passLine,
  boxrule=0.5pt,
  arc=5pt,
  left=5pt,
  right=5pt,
  top=1pt,
  bottom=1pt,
  fontupper=\sffamily\bfseries\tiny\color{passText}
}

\newtcbox{\failbox}{
  on line,
  colback=failFill,
  colframe=failLine,
  boxrule=0.5pt,
  arc=5pt,
  left=5pt,
  right=5pt,
  top=1pt,
  bottom=1pt,
  fontupper=\sffamily\bfseries\tiny\color{failText}
}

\newcommand{\testspassed}[1]{%
  \passbox{\ding{51}\ #1 TESTS PASSED}%
}

\newcommand{\testsfailed}[1]{%
  \failbox{\ding{55}\ #1 TESTS FAILED}%
}

\newcommand{\gridlabel}[1]{%
  {\sffamily\bfseries\tiny\MakeUppercase{#1}}%
}

\tcbset{
  colhead/.style={
    overlay app={
      \node[anchor=south, inner sep=0pt]
      at ([yshift=4pt]frame.north)
      {\gridlabel{#1}};
    }
  },
  rowlabel/.style={
    overlay app={
      \node[
        anchor=east,
        inner sep=0pt,
        align=right,
        font=\sffamily\bfseries\tiny
      ]
      at ([xshift=-5pt]frame.west)
      {#1};
    }
  },
}

\newtcblisting{casecode}[2][]{
  enhanced,
  listing and comment,
  listing options={style=casecode},
  comment={\centering #2\par},
  colback=lightgrey,
  colframe=lightgrey,
  boxrule=0pt,
  arc=4pt,
  left=2pt,
  right=2pt,
  top=3pt,
  bottom=2pt,
  segmentation hidden,
  colbacklower=lightgrey!60!gray!25,
  middle=3pt,
  space to upper,
  #1
}

\begin{document}

\title{Do Modifications Made by Coding Agents Make Subsequent Software Maintenance Harder?}


\author{Ruwei Pan}
\authornote{Both authors contributed equally to this research.}
\email{panruwei666@gmail.com}
\affiliation{%
  \institution{Peking University}
  \country{China}
}

\author{John Zhang}
\authornotemark[1]
\email{john.zhang959@gmail.com}
\affiliation{%
  \institution{University of New South Wales}
  \country{Australia}
}

\author{Linhao Wu}
\email{wulinhao26@stu.pku.edu.cn}
\affiliation{%
  \institution{Peking University}
  \country{China}
}

\author{Yakun Zhang}
\email{zhangyk@hit.edu.cn}
\affiliation{%
  \institution{Harbin Institute of Technology (Shenzhen)}
  \country{China}
}

\author{Hongyu Zhang}
\email{hongyujohn@gmail.com}
\affiliation{%
  \institution{Chongqing University}
  \country{China}
}

\author{Lu Zhang}
\authornote{Corresponding author.}
\email{zhanglucs@pku.edu.cn}
\affiliation{%
  \institution{Peking University}
  \country{China}
}

\renewcommand{\shortauthors}{Pan et al.}

\begin{abstract}

Coding agents are increasingly used to maintain repositories over time, and each change an agent makes becomes the starting point of its next maintenance task.
Yet benchmarks such as SWE-bench start every task from the developers' code, and sequential benchmarks such as ChainSWE cannot separate a later task's difficulty from the code the agent inherited.
This leaves a practical question open: does agent-written code itself hinder later maintenance, or does the cost come from earlier tests that the inherited code already fails?
We introduce \textbf{SWE-Maintain}, 100 three-event maintenance episodes reconstructed from 50 real-world Python repositories, with hidden tests certified to hold across each sequence.
Each later event is run twice with the same task, agent, and tests: once from the developers' preceding patches (\textbf{Developer-History}, as in SWE-bench) and once from the agent's own preceding changes (\textbf{Agent-History}).
The paired difference measures the cost of building on the agent's own history.
Across five coding systems and two backbone models (1,996 paired events), building on its own history lowers an agent's later-task resolution in all ten combinations, by 5.6 percentage points on average.
When all earlier tests must also pass, the gap grows to 14.0 points and is concentrated in inherited failures: two-thirds of Agent-History starting repositories already fail earlier tests, and only 1.9\% of events built on them end with all accumulated tests passing.
Once the inherited repository passes all earlier tests and the next task is not already solved, the disadvantage disappears ($+2.7$ points, 95\% CI [$-0.5$, $+6.1$]).
For subsequent agents, the debt that coding agents leave behind thus lies mainly in earlier tests left failing rather than in harder-to-maintain code.
This suggests that evaluations should measure agents from their own history, and that agents should prioritize preserving earlier behavior across changes.

\end{abstract}



\maketitle

\section{Introduction}

Coding agents can now perform repository-level software maintenance tasks end to end, including locating relevant code, modifying multiple files, and validating their changes through execution and testing~\cite{jimenez2024swe,yang2024swe,zhang2024autocoderover,wang2025openhands,chen2025locagent}. As these agents are increasingly used for real-world software development, their role is shifting from solving isolated tasks to repeatedly maintaining software repositories over time. In this setting, an agent’s modification does not disappear after the current task is completed; it becomes part of the repository state on which subsequent maintenance must build. Consequently, evaluating each task independently from developer-written code, as in commonly used benchmarks such as SWE-bench~\cite{jimenez2024swe}, captures only an agent’s ability to solve individual tasks but not its ability to solve a series of maintenance tasks over time. 
Recent sequential benchmarks carry an agent's changes forward~\cite{deng2026swe,jin2026chainswe}, but they still judge each task only by whether its own test cases pass, not by whether an earlier change made later tasks harder.

Whether an agent's earlier changes make later maintenance harder is a question of \emph{technical debt}, the cost that a change satisfying an immediate requirement can impose on future work~\cite{cunningham1992wycash,ramasubbu2018integrating,guo2025understanding}.
Such downstream maintenance cost can arise for 
different reasons.
For coding agents, this downstream cost can arise in two fundamentally different ways.
The first is \emph{code debt}~\cite{alves2016identification}, in which the agent's code is itself harder to build on, for example because it touches more files than needed, adds dependencies, or organizes functionality differently from the developers, even when it passes the current task's tests~\cite{xiong2018identifying,xin2017identifying,cai2025understanding}.
The second is \emph{inherited test failures}, tests of earlier events that already fail in the code a later task starts from, which the next agent inherits without being asked to fix them.
The two sources call for different remedies, improving the code that agents write in the first case and keeping test failures from carrying forward in the second.
These considerations lead to our central question.
\textbf{When a coding agent builds on its own earlier changes, does subsequent maintenance become harder, and if so, does the cost come from code debt or from inherited test failures?}

Existing benchmarks cannot answer this question, for two reasons.
First, benchmarks that start every task from the developers' code never expose an agent to its own earlier changes, so they cannot observe any inherited cost.
Second, in sequential benchmarks, the outcome of a later task reflects both the task's own difficulty and the code inherited from earlier agent changes, and a single trajectory cannot separate the two.
Separating them requires observing the same later task under both kinds of history.

\begin{figure}[t]
    \centering
    \includegraphics[width=0.9\columnwidth]{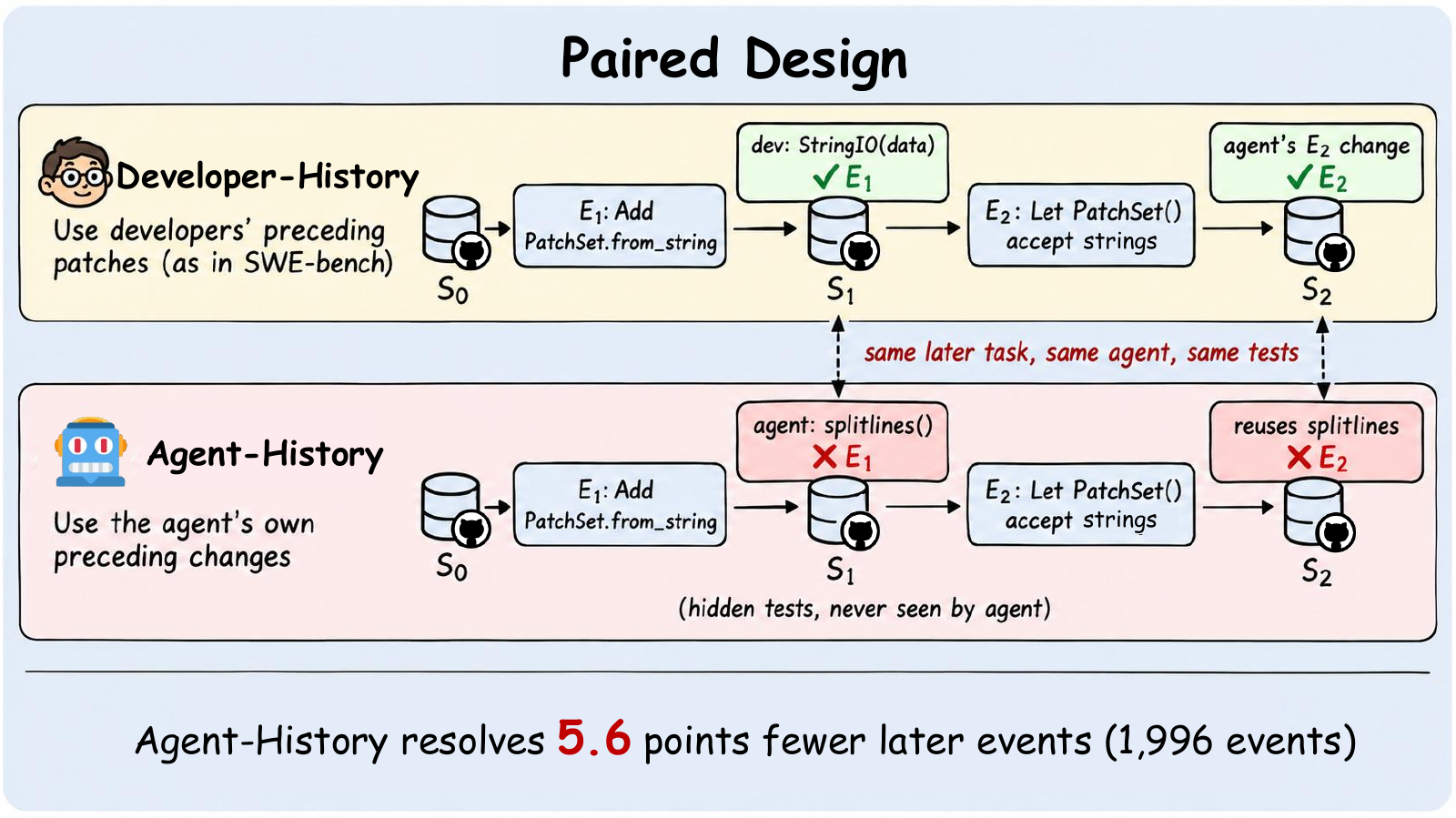}
    \caption{
    The same later task from two starting points (\texttt{python\_unidiff}, Claude Code with DeepSeek-V4-Flash).
    At $E_1$, the agent's change splits the input into lines and fails the event's hidden tests, which the agent never sees.
    At $E_2$, the same agent solves the next task when it starts from the developers' $E_1$ patch (Developer-History), but fails when it starts from its own $E_1$ change (Agent-History), because it reuses the defective logic.
    Over all 1{,}996 later events, agents resolve 5.6 points fewer tasks under Agent-History than under Developer-History. 
    }
    \label{fig:intro_design}
    \vspace{-10px}
\end{figure}

Our key observation is that the first kind of benchmark already supplies one side of this comparison.
Running a task from the developers' preceding patches, as SWE-bench does, gives a reference outcome, and running the same task from the agent's own preceding changes gives its outcome under continuous maintenance.
With the task, agent, and tests held fixed, the difference between the two outcomes measures the cost of building on the agent's own history, relative to the developers' original patches.
We realize this design in \textbf{SWE-Maintain}, which contains 100 three-event maintenance episodes reconstructed from 50 real-world Python repositories.
We certify each episode by replaying the developers' patches in order and requiring all tests from earlier events to keep passing, so that a later failure of those tests reflects a change made during evaluation rather than a conflict already present in the original history.
Each later event is run in two paired settings, \textbf{Developer-History}, which starts from the developers' patches for all preceding events, and \textbf{Agent-History}, which starts from the agent's own changes for those events.
Figure~\ref{fig:intro_design} illustrates the design on a real episode from SWE-Maintain.
At $E_1$, the agent's change fails the event's hidden tests; at $E_2$, the same agent solves the next task from the developers' code but fails it from its own, because it reuses its earlier defective logic.
We quantify this effect as the paired difference in resolution rate between Agent-History and Developer-History on the same later events, so a negative value means that agents do worse on their own history.

Using SWE-Maintain, we study five coding systems with two backbone models, covering 1,996 paired later events, and ask three research questions.
\textbf{RQ1} measures the cost, comparing later-task resolution, preservation of earlier tests, and effort between Agent-History and Developer-History.
\textbf{RQ2} locates the source of this cost, comparing events whose inherited code already fails earlier tests with those whose inherited code passes them, and removing future tasks that earlier changes had already solved.
\textbf{RQ3} explores, among inherited code that passes all earlier tests, which properties of the agent's earlier changes are associated with helping or hindering the next task.

Our study yields three main findings.
(1) Building on its own history lowers an agent's later-task resolution in all ten system--model combinations, by 5.6 percentage points (95\% CI [$-7.2$, $-4.1$]), whereas at the first event, where both settings start from identical code, they differ by only $+1.8$ points.
(2) When all earlier tests must also pass, the gap grows to 14.0 points and is concentrated in inherited test failures.
Two-thirds of Agent-History starting repositories already fail earlier tests, only 1.9\% of the events built on them end with all accumulated tests passing, and even the new task alone is solved 7.6 points less often there than from developer history.
(3) Once the inherited code passes all earlier tests and the next task is not already solved, the agent does as well as from developer history ($+2.7$ points, 95\% CI [$-0.5$, $+6.1$]), which rules out a disadvantage larger than half a point.
Within such code, the clearest pattern in our exploratory analysis is that earlier changes overlapping with functions the next task later modifies are associated with more successes under Agent-History alone and with lower token usage.
These findings point to the second source.
For subsequent agents, the cost of building on agent-written code lies mainly in inherited test failures rather than in code debt.
This result has two implications.
Benchmarks that start every task from developers' code overstate how well agents sustain maintenance, so evaluations should also run agents from their own history.
Agents, in turn, should ensure that all earlier tests still pass before considering a change complete.

\setlist[itemize]{leftmargin=1em, labelsep=0.6em, itemsep=2pt}

In summary, the main contributions of this work are as follows:

\begin{itemize}
    \item A paired evaluation design that isolates the cost of building on an agent's own history by running each later task from both the developers' and the agent's preceding changes, where the developer side corresponds to standard benchmark evaluation.

    \item \textbf{SWE-Maintain}, a benchmark for evaluating downstream maintenance in agent-based software development, which consists of 100 behaviorally certified three-event episodes from 50 real-world Python repositories, released with its evaluation harness and all agent trajectories.

    \item An empirical study of five coding systems and two backbone models showing that the cost of continuous agent maintenance is concentrated in inherited test failures, while agent-written code that passes all earlier tests imposes no measurable disadvantage on subsequent agents.
\end{itemize}

\section{SWE-Maintain}
\label{sec:swe_maintain}

\subsection{Overview}

\begin{figure*}[t]
    \centering
    \includegraphics[width=\textwidth]{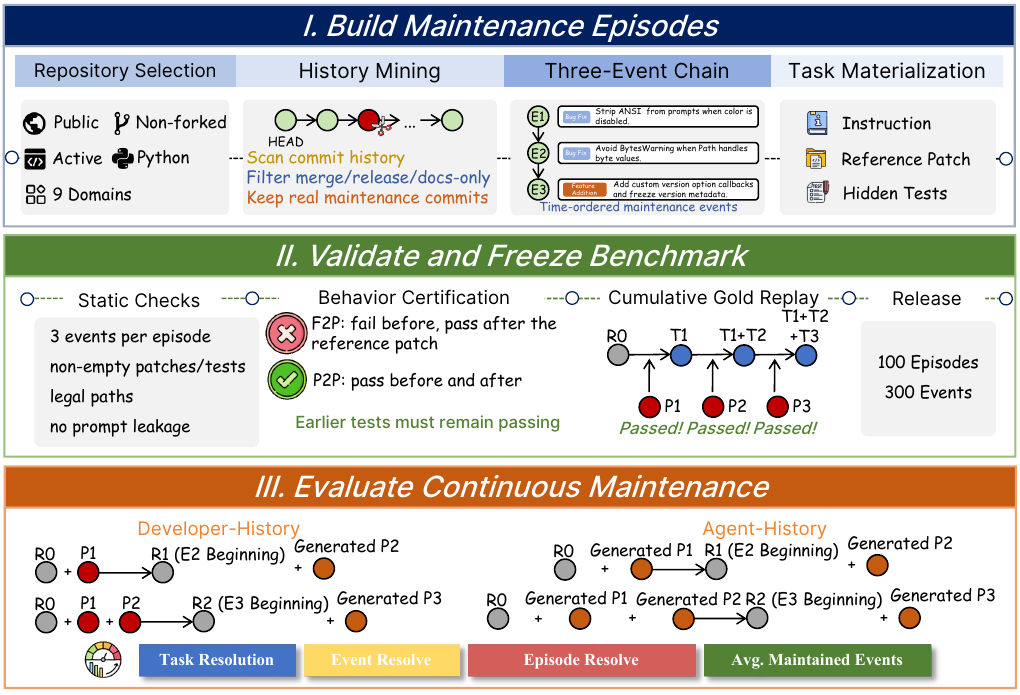}
    \caption{Overview of SWE-Maintain. Maintenance episode construction and benchmark validation produce certified three-event episodes. The continuous maintenance evaluation protocol runs each later event from the developers' preceding patches (Developer-History) and from the agent's own preceding changes (Agent-History).}
    \label{fig:swe_maintain_overview}
    \vspace{-10px}
\end{figure*}

SWE-Maintain is a benchmark for measuring how an agent's earlier changes affect its later maintenance.
It contains 100 episodes from 50 real-world Python repositories; each episode is a sequence of three maintenance events taken from a repository's history.
The benchmark evaluates every later event twice, once on top of the developers' history (\emph{Developer-History}) and once on top of the agent's own (\emph{Agent-History}). 
Figure~\ref{fig:swe_maintain_overview} shows the three stages that produce and use these episodes.
\emph{Maintenance episode construction} (Section~\ref{sec:construction}) mines events from commit histories and turns each into a task with an instruction, a developer reference patch, and hidden tests.
\emph{Benchmark validation and freezing} (Section~\ref{sec:certification}) checks every event on its own and every episode as a replayed sequence, then releases the certified episodes.
The \emph{continuous maintenance evaluation protocol} (Section~\ref{sec:paired_protocol}) runs each later event under Developer-History and Agent-History.

\subsection{Maintenance Episode Construction}
\label{sec:construction}

Each episode must preserve the order of real maintenance changes while keeping every event independently executable and testable.
As shown in Fig.~\ref{fig:swe_maintain_overview}, the construction process consists of repository selection, history mining, three-event chain construction, and task materialization.

\textbf{Repository Selection.}
We collect candidate repositories from GitHub through its search API, keeping public, non-fork, and non-archived repositories whose primary language is Python.
Because stars and forks reflect community adoption, we require at least 100 stars and 10 forks, so that each repository has users whose needs drive ongoing maintenance.
We further require at least 50 commits on the default branch and at least one commit in the 12 months before collection, so that the history contains enough changes to mine and the project is still maintained.
To keep environment setup and test execution tractable, we exclude repositories with more than 150K lines of Python code.
Finally, we keep only repositories with a \texttt{pytest}- or \texttt{unittest}-based test suite that can be installed from the repository's dependency specification and run locally without inaccessible external services or private infrastructure.
SWE-Maintain retains the 50 repositories from which at least one episode passes the certification in Section~\ref{sec:certification}.

\textbf{History Mining.}
For each selected repository, we inspect its commit history and identify commits that represent concrete and independently testable maintenance changes.
We consider common maintenance activities, including bug fixes, feature additions, behavior or interface modifications, robustness improvements, and compatibility-related changes.
We exclude merge commits, release-related commits, documentation-only changes, formatting changes, dependency-only updates, pure refactorings without observable behavioral changes, and commits that mix multiple unrelated modifications.
For each remaining candidate, we recover the repository state immediately before the change and retain the corresponding developer patch as a reference patch.

\textbf{Three-Event Chain.}
We organize candidate events into three-event chains,
$E_1 \rightarrow E_2 \rightarrow E_3$,
following their original temporal order in the repository history.
Events need not be adjacent in the original history.
When commits between two selected events are not themselves selected, we omit them and apply the selected developer patches in order; this holds for 139 of the 200 later events, with 1 to 316 omitted commits.
We use three events because this is the shortest sequence that exposes both one-step and accumulated downstream effects.
$E_2$ reflects the effect of the modification introduced at $E_1$, while $E_3$ can reflect the combined effect of the modifications introduced at $E_1$ and $E_2$.
We retain only chains whose developer patches can be replayed in order and pass the certification in Section~\ref{sec:certification}.
Each later event is therefore performed on code that already contains the changes of the events before it, as in the original history.

\textbf{Task Materialization.}
Each selected historical change is converted into an executable maintenance task consisting of a maintenance instruction, a developer reference patch, and hidden tests.
We reconstruct the maintenance instruction from the original commit message together with its associated pull request or issue when available.
Instructions are reviewed in two stages.
First, two independent LLM reviewers check that every hidden test traces to a requirement stated in the instruction and that the instruction reveals neither the developer patch nor the hidden tests; they agreed on all 300 events.
Second, two human reviewers who are not authors of this paper independently inspect every instruction against four criteria: it states the required observable behavior, it is consistent with the original commit and its linked pull request or issue, it does not prescribe the developer's implementation, and every hidden test checks behavior the instruction requires.
Disagreements are resolved by discussion, and no instruction required rewriting.
The developer reference patch is used only for task validation and reconstruction of the developer-produced reference state.
Hidden tests specify the required behavior of each event and are never exposed to the evaluated approach.
For 192 events, they are the developers' own test changes from the source commit.
The remaining 108 events come from commits that either change no tests or whose test changes do not distinguish a correct implementation from the code before the change.
For these events, we write tests from the instruction alone, without consulting the developer patch, and certify them in the same way as upstream tests.
The resulting three tasks form one maintenance episode in SWE-Maintain.

\subsection{Benchmark Validation and Freezing}
\label{sec:certification}

A commit from a repository's history does not by itself make a usable evaluation event.
Its tests may give inconsistent results across runs or depend on an environment that cannot be reproduced, they may not detect the change the commit makes, and a later commit in the same episode may intentionally change behavior that an earlier event's tests check.
We address these problems with three checks, shown in the validation stage of Fig.~\ref{fig:swe_maintain_overview}.
\emph{Static checks} discard events with missing patches or tests, inconsistent test results, unreproducible dependencies, or instructions that expose the solution.
\emph{Behavior certification} checks each event on its own: its hidden tests must fail before the developer patch and pass after it, and tests that passed before the patch must keep passing.
\emph{Cumulative gold replay} checks each episode as a sequence: after each developer patch is applied in order, the tests of the current event and of all earlier events must pass, except tests whose behavior a later developer change intentionally redefines.
We call an episode \emph{certified} when it passes all three checks, and we freeze and release only certified episodes.
Certification lets us attribute regressions to the agent.
In a certified episode, the developers' own patches pass the tests of all earlier events at every step, apart from tests that a later developer change intentionally redefines.
Therefore, if an earlier event's test fails under Agent-History, the failure cannot come from the benchmark itself; it must come from code the agent wrote during the episode.

\textbf{Static Checks.}
We first perform basic validity checks on each candidate episode.
Every event must have a non-empty developer reference patch and executable hidden tests.
We verify that all modified files belong to valid repository paths and that neither the maintenance instruction nor the evaluation environment exposes the developer patch, hidden tests, or other solution-specific information.
We also execute the tests repeatedly and discard tasks whose results are inconsistent across runs, whose dependencies cannot be reproduced, or that rely on inaccessible external services.

\textbf{Behavior Certification.}
For each maintenance event $E_i$, we verify that the developer reference patch introduces the required behavior while preserving existing tested functionality.
Let $S_{i-1}$ denote the repository state before $E_i$, and let $P_i$ denote its developer reference patch, such that $S_i=P_i(S_{i-1})$.
The Fail-to-Pass tests for $E_i$ must fail before applying $P_i$ and pass afterward:
\begin{equation}
T_i^{\mathrm{F2P}}(S_{i-1})=\mathrm{Fail},
\qquad
T_i^{\mathrm{F2P}}(S_i)=\mathrm{Pass}.
\end{equation}
The Pass-to-Pass tests must remain passing across the change:
\begin{equation}
T_i^{\mathrm{P2P}}(S_{i-1})=
T_i^{\mathrm{P2P}}(S_i)=\mathrm{Pass}.
\end{equation}
Each hidden test is also traced to a requirement of the instruction, as described in Section~\ref{sec:construction}.
These checks establish that each event has an observable behavioral change and can be evaluated independently.

\textbf{Cumulative Gold Replay.}
Behavior certification alone does not guarantee that a sequence of historical changes forms a consistent maintenance episode.
A later developer change may invalidate behavior introduced by an earlier event.
We therefore replay the three developer reference patches sequentially from the initial repository state:
\begin{equation}
S_0
\xrightarrow{P_1}
S_1
\xrightarrow{P_2}
S_2
\xrightarrow{P_3}
S_3.
\end{equation}
After each patch, we execute the tests for the current event and all preceding events.
When a later developer change intentionally redefines behavior that an earlier event tested, we retire the superseded tests from that point on, which affects six $E_3$ events.
Let $T_i$ denote the tests of $E_i$ that are not superseded by a later event.
We require
\begin{equation}
S_1 \models T_1,\qquad
S_2 \models T_1 \cup T_2,\qquad
S_3 \models T_1 \cup T_2 \cup T_3.
\end{equation}
An episode is retained only when all three developer patches can be replayed in order and all previously established tested behaviors remain valid after each subsequent change.
We also check that later instructions do not depend on implementation-specific details of preceding developer patches, so that a later task is not defined around the particular implementation chosen earlier in the developer history.

\subsection{Benchmark Statistics}
\label{sec:benchmark_stats}

The release step of Fig.~\ref{fig:swe_maintain_overview} freezes 100 episodes (300 events) from 50 repositories; Table~\ref{tab:benchmark_stats} summarizes them.
The 50 repositories span nine application domains, led by developer tools, command-line tools, HTTP clients, and general-purpose libraries, and range from 160 to 128,632 lines of Python (median 1,744).
Events cover nine maintenance intents: bug fixes, feature additions, and behavior changes account for 80\% of them, and about one third span several modules or change a public API.
Reference patches are focused, with a median of one changed non-test file and 17 changed lines, which reflects the atomic commits from which events are mined and keeps each event attributable to a single maintenance intent.
Most sequences are connected: 145 of the 200 later events extend, modify, or depend on behavior introduced earlier in their episode, and 122 (61\%) change a file that an earlier developer patch in the same episode also changed.
The remaining 55 later events are functionally independent of earlier ones but still start from code that contains their changes.
Intent and relation labels were assigned by two independent LLM annotators with adjudication, using the definitions released with the benchmark. 

\begin{table}[t]
\centering
\caption{\textbf{Statistics of SWE-Maintain.} Distributions are given as median [interquartile range].}
\label{tab:benchmark_stats}
\small
\setlength{\tabcolsep}{3pt}
\renewcommand{\arraystretch}{1.1}
\begin{adjustbox}{max width=\columnwidth}
\begin{tabular}{@{}lr@{}}
\toprule
\multicolumn{2}{@{}l}{\textbf{Repositories}} \\
Repositories / episodes / events & 50 / 100 / 300 \\
Domains & 9 \\
\quad developer tools / CLI tools / HTTP clients & 16 / 8 / 7 \\
\quad libraries / data processing / infrastructure / frameworks & 7 / 5 / 4 / 3 \\
Python LOC per repository & 1,744 [977, 5,208] \\
Python modules per repository & 14 [10, 33] \\
\midrule
\multicolumn{2}{@{}l}{\textbf{Events}} \\
Bug fix / feature addition / behavior change & 90 / 81 / 69 \\
Configuration or schema / API compatibility & 17 / 14 \\
Error handling / dependency or build / other & 12 / 10 / 7 \\
Cross-module events / public-API events & 101 / 99 \\
Non-test files changed per reference patch & 1 [1, 2] \\
Non-test lines changed per reference patch & 17 [6, 39] \\
Hidden tests from upstream / written from instruction & 192 / 108 \\
Instruction length (words) & 24 [22, 28] \\
\midrule
\multicolumn{2}{@{}l}{\textbf{Sequences (200 later events)}} \\
Extend / modify / depend on earlier behavior & 106 / 14 / 25 \\
Independent of earlier events & 55 \\
Change a file changed by an earlier event & 122 \\
With omitted intermediate commits & 139 \\
\bottomrule
\end{tabular}
\end{adjustbox}
\vspace{-10px}
\end{table}

\subsection{Continuous Maintenance Evaluation Protocol}
\label{sec:paired_protocol}

Most later events in SWE-Maintain build on code that earlier events changed: 145 of the 200 later events extend, modify, or depend on behavior introduced earlier in their episode (Section~\ref{sec:benchmark_stats}).
How well an agent performs such an event may depend on who wrote those earlier changes.
Standard benchmarks always start a task from the developers' code, whereas an agent used for continuous maintenance starts from the changes it made itself.
We therefore evaluate each later event in two settings that differ only in the code it starts from. 
Let $A_i$ denote the change an approach makes at $E_i$.
\textbf{Developer-History} starts $E_i$ from $S_{i-1}$, the code after the developers' patches for all preceding events.
\textbf{Agent-History} starts $E_i$ from $\hat{S}_{i-1}$, where $\hat{S}_0 = S_0$ and $\hat{S}_i = A_i(\hat{S}_{i-1})$; if the approach produces no valid change, $\hat{S}_i = \hat{S}_{i-1}$.
At $E_2$, the comparison therefore reflects the agent's change at $E_1$, and at $E_3$ the combined changes at $E_1$ and $E_2$.

Within each approach--model combination, the two settings use the same instruction, backbone model, execution environment, stopping rule, execution limits, and hidden tests.
Each event runs in a fresh session, so no conversation history or temporary agent state carries over between events; only the repository itself is carried forward in Agent-History.

For a later event, let $Y^{D}$ and $Y^{A}$ be its binary outcomes under the two settings.
Our estimand is the paired difference $\Delta = \mathbb{E}[Y^{A} - Y^{D}]$, averaged over events.
Each setting is run once per event, so individual pairs can differ by chance.
Because both settings start $E_1$ from the same code $S_0$, the $E_1$ difference measures this run-to-run variation directly, and we use it as a reference for the later-event differences.
We report $\Delta$ with repository-clustered bootstrap intervals (Section~\ref{sec:stats}).

\section{Evaluating Coding Agents on SWE-Maintain}

SWE-Maintain provides the episodes and the paired protocol. The object of study are the coding agents that maintain a repository over a sequence of changes.
We evaluate five coding systems with two backbone models on SWE-Maintain (Section~\ref{sec:swe_maintain}), covering 1{,}996 paired later events.
Our study follows the central question of Section~1: when a coding agent builds on its own earlier changes, does subsequent maintenance become harder, and if so, does the cost come from code debt or from inherited test failures?
We address it with three research questions: the first measures the cost, the second separates its two candidate sources, and the third explores which properties of an earlier change predict whether it helps or hinders the next task. 

\textbf{RQ1. What does building on its own history cost a coding agent in subsequent maintenance?}
We run every later event under Developer-History and Agent-History and compare whether the new task is solved, whether the tests of earlier events still pass, and how much effort the agent spends.

\textbf{RQ2. Does the cost come from inherited test failures or from the agent-written code itself?}
If agent-written code carried code debt, the cost would persist even when that code passes all earlier tests.
If the cost came from inherited test failures, it would concentrate where those tests fail.
We therefore split later events by whether their Agent-History starting code passes all earlier tests, and we further remove events whose task the earlier changes had already solved.

\textbf{RQ3. Within inherited code that passes all earlier tests, which properties of the agent's earlier changes are associated with helping or hindering the next task?}
This exploratory analysis relates six properties of each earlier agent change, measured relative to the developers' change for the same event, to later success and effort.

\subsection{Metrics}

We measure outcomes at the event and episode levels.
At the event level, we separate whether the new task is solved from whether the work of earlier events survives, because a change can satisfy the first while breaking the second.

\ding{182} \textbf{Task Resolution Rate (TR).}
TR is the percentage of events whose resulting repository passes the hidden tests of the current event, regardless of earlier tests.
We report TR separately for $E_1$, $E_2$, and $E_3$.
We also report \textbf{Later TR}, computed over $E_2$ and $E_3$, to summarize current-task resolution on events whose starting states depend on preceding modifications.

\ding{183} \textbf{Event Resolve Rate (ER).}
ER is the percentage of events whose resulting repository passes both the hidden tests of the current event and all tests $T_1, \ldots, T_{i-1}$ of earlier events.
It is the outcome a maintainer cares about: the request is done and nothing that earlier work established is broken.
Under Developer-History, the starting code passes all earlier tests by certification (Section~\ref{sec:certification}), so any ER failure there comes from the current change alone.
We report \textbf{Later ER} over $E_2$ and $E_3$.

\ding{184} \textbf{Episode Resolve Rate (EpR) and Average Maintained Events (AME).}
EpR is the percentage of Agent-History episodes in which all three events are resolved under ER, and AME is the mean number of consecutive ER-resolved events from the start of an episode, ranging from 0 to 3.
Both describe a continuous trajectory and are therefore reported for Agent-History only.

\ding{185} \textbf{Maintenance Effort.}
We measure effort by consumed tokens and model requests.
Because failed runs stop at different points, we compare effort only on events resolved (ER) under both settings, using the per-event log ratio $\log\frac{x^{A}+1}{x^{D}+1}$, where $x^{A}$ and $x^{D}$ are the effort under Agent-History and Developer-History.

\subsection{Statistical Analysis}
\label{sec:stats}

All differences are Agent-History minus Developer-History, so a negative $\Delta$TR or $\Delta$ER is a cost of building on the agent's own history.
We estimate 95\% confidence intervals with a repository-clustered bootstrap stratified by approach and model (10,000 replicates), which accounts for repeated events within a repository.
We also report the discordant pairs, events resolved under only one setting, and test them with an exact McNemar test.
Because both settings start $E_1$ from the same code, the $E_1$ difference serves as a reference for run-to-run variation (Section~\ref{sec:paired_protocol}).
For effort, we report the median log ratio with a clustered bootstrap interval.

\subsection{Studied Coding Agents}
\label{sec:studied_coders}
We study five systems for repository-level software engineering that differ in how they interact with a repository.

\begin{itemize}

    \item \textbf{Mini-SWE-Agent~\cite{yang2024swe}} is a lightweight code agent designed for real-world software engineering tasks. It uses a simple interaction loop to inspect repositories, execute commands, and modify code, relying largely on the capability of the backbone model.

    \item \textbf{Agentless~\cite{xia2025demystifying}} solves tasks through a fixed pipeline of fault localization, patch generation, and patch validation rather than an autonomous loop. It lets us test whether the effect depends on agentic interaction.

    \item \textbf{SE-Agent~\cite{guo2026se}} improves code-agent problem solving by learning from multiple previous attempts. It revises and combines previous interaction trajectories to explore better solutions for software engineering tasks.

    \item \textbf{OpenHands~\cite{wang2025openhands}} is a general-purpose platform for autonomous software development. It allows code agents to inspect and modify repositories, execute commands and tests, and revise their solutions based on execution feedback.

    \item \textbf{Claude Code~\cite{anthropic2025claudecode}} is a code agent developed by Anthropic for real-world software development. It can inspect codebases, modify files, execute commands and tests, and revise its implementation based on execution feedback.

\end{itemize}

\subsection{Implementation Details}

We evaluate all applicable approaches with two backbone models, \textbf{GPT-5.6-Luna} and \textbf{DeepSeek-V4-Flash-0731}, to examine whether the observed downstream patterns are consistent across different backbone models.
For each approach--model combination, Developer-History and Agent-History use the same model version and configuration.
We reproduce each approach using its official implementation and prompt when available, and otherwise follow the settings recommended by its authors.

Both settings are run independently, once per event, and all patches are evaluated with the same hidden tests and timeout.
Agentless with DeepSeek-V4-Flash-0731 has paired runs for 98 of the 100 episodes, which gives 1,996 paired later events in total.
The exact model parameters, framework versions, stopping conditions, and execution limits are reported in our repository.

\section{Evaluation Results}

\subsection{RQ1. The Maintenance Cost of Building on One's Own History}
\label{sec:rq1}

\begin{table*}[t]
\centering
\caption{\textbf{Subsequent maintenance under Developer-History (Dev.) and Agent-History (Agent).}
TR, ER, and EpR are percentages; $\Delta$ is Agent minus Dev.\ in percentage points.
$E_1$ starts from identical code in both settings and serves as a reference.
Later TR and Later ER are computed over $E_2$ and $E_3$.
EpR and AME describe continuous trajectories and are reported for Agent-History only.
For Agentless with DeepSeek-V4-Flash-0731, Later ER uses its 98 paired episodes.}
\label{tab:rq1_main}

\footnotesize
\setlength{\tabcolsep}{3.2pt}
\renewcommand{\arraystretch}{1.12}

\begin{adjustbox}{max width=\textwidth}
\begin{tabular}{@{}llccccccccc@{}}
\toprule
\multirow{2}{*}{\textbf{Approach}} &
\multirow{2}{*}{\textbf{Setting}} &
\multicolumn{5}{c}{\textbf{Task Resolution}} &
\multicolumn{2}{c}{\textbf{Event Resolve}} &
\multicolumn{2}{c}{\textbf{Episode}} \\
\cmidrule(lr){3-7}
\cmidrule(lr){8-9}
\cmidrule(lr){10-11}
& &
\textbf{$E_1$} & \textbf{$E_2$} & \textbf{$E_3$} & \textbf{Later TR} & \textbf{$\Delta$TR} &
\textbf{Later ER} & \textbf{$\Delta$ER} &
\textbf{EpR} & \textbf{AME} \\
\midrule
\rowcolor{lightgrey}
\multicolumn{11}{c}{\textbf{GPT-5.6-Luna}} \\
\midrule
\multirow{2}{*}{\textbf{Mini-SWE-Agent}}
& Dev.  & 48.0 & 46.0 & 63.0 & 54.5 & & 42.0 & & -- & -- \\
& Agent & 49.0 & 45.0 & 51.0 & 48.0 & \tblimppos{$-$6.5} & 21.5 & \tblimppos{$-$20.5} & 18.0 & 0.91 \\
\addlinespace[2pt]
\multirow{2}{*}{\textbf{Agentless}}
& Dev.  & 26.0 & 26.0 & 36.0 & 31.0 & & 20.5 & & -- & -- \\
& Agent & 30.0 & 20.0 & 32.0 & 26.0 & \tblimppos{$-$5.0} & 9.5 & \tblimppos{$-$11.0} & 7.0 & 0.48 \\
\addlinespace[2pt]
\multirow{2}{*}{\textbf{SE-Agent}}
& Dev.  & 48.0 & 48.0 & 62.0 & 55.0 & & 41.5 & & -- & -- \\
& Agent & 48.0 & 40.0 & 53.0 & 46.5 & \tblimppos{$-$8.5} & 20.5 & \tblimppos{$-$21.0} & 17.0 & 0.88 \\
\addlinespace[2pt]
\multirow{2}{*}{\textbf{OpenHands}}
& Dev.  & 50.0 & 51.0 & 61.0 & 56.0 & & 44.5 & & -- & -- \\
& Agent & 52.0 & 46.0 & 56.0 & 51.0 & \tblimppos{$-$5.0} & 29.0 & \tblimppos{$-$15.5} & 24.0 & 1.05 \\
\addlinespace[2pt]
\multirow{2}{*}{\textbf{Claude Code}}
& Dev.  & 52.0 & 47.0 & 63.0 & 55.0 & & 41.0 & & -- & -- \\
& Agent & 53.0 & 44.0 & 55.0 & 49.5 & \tblimppos{$-$5.5} & 24.0 & \tblimppos{$-$17.0} & 20.0 & 1.00 \\
\midrule
\rowcolor{lightgrey}
\multicolumn{11}{c}{\textbf{DeepSeek-V4-Flash-0731}} \\
\midrule
\multirow{2}{*}{\textbf{Mini-SWE-Agent}}
& Dev.  & 43.0 & 43.0 & 60.0 & 51.5 & & 37.5 & & -- & -- \\
& Agent & 50.0 & 42.0 & 49.0 & 45.5 & \tblimppos{$-$6.0} & 24.0 & \tblimppos{$-$13.5} & 17.0 & 0.95 \\
\addlinespace[2pt]
\multirow{2}{*}{\textbf{Agentless}}
& Dev.  & 11.0 & 6.0 & 16.0 & 11.0 & & 4.1 & & -- & -- \\
& Agent & 14.0 & 4.0 & 11.0 & 7.5 & \tblimppos{$-$3.5} & 1.0 & \tblimppos{$-$3.1} & 0.0 & 0.15 \\
\addlinespace[2pt]
\multirow{2}{*}{\textbf{SE-Agent}}
& Dev.  & 41.0 & 36.0 & 50.0 & 43.0 & & 29.5 & & -- & -- \\
& Agent & 41.0 & 36.0 & 49.0 & 42.5 & \tblimppos{$-$0.5} & 18.0 & \tblimppos{$-$11.5} & 13.0 & 0.73 \\
\addlinespace[2pt]
\multirow{2}{*}{\textbf{OpenHands}}
& Dev.  & 46.0 & 43.0 & 53.0 & 48.0 & & 37.0 & & -- & -- \\
& Agent & 49.0 & 37.0 & 47.0 & 42.0 & \tblimppos{$-$6.0} & 24.5 & \tblimppos{$-$12.5} & 17.0 & 0.92 \\
\addlinespace[2pt]
\multirow{2}{*}{\textbf{Claude Code}}
& Dev.  & 50.0 & 45.0 & 57.0 & 51.0 & & 38.0 & & -- & -- \\
& Agent & 46.0 & 37.0 & 46.0 & 41.5 & \tblimppos{$-$9.5} & 23.5 & \tblimppos{$-$14.5} & 19.0 & 0.89 \\
\bottomrule
\end{tabular}
\end{adjustbox}
\vspace{-10px}
\end{table*}

\begin{figure}[t]
  \centering
  \includegraphics[width=\columnwidth]{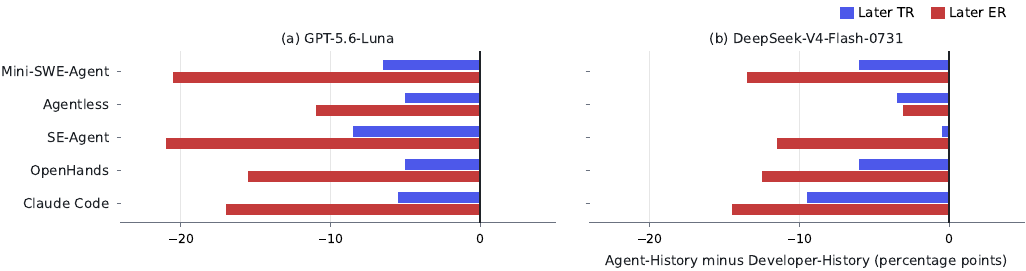}
  \caption{\textbf{Cost of building on the agent's own history in RQ1.}
  Paired difference between Agent-History and Developer-History for every approach--model combination, in percentage points.
  Negative values indicate a lower rate under Agent-History.
  Later ER additionally requires all tests from earlier events to pass, and its cost exceeds the Later TR cost in nine of the ten combinations.
  Values are from Table~\ref{tab:rq1_main}.}
  \label{fig:rq1_cost}
  \vspace{-10px}
\end{figure}

Table~\ref{tab:rq1_main} reports both settings for every approach--model combination.
We define \emph{maintenance cost} as the performance loss when maintaining Agent-History rather than Developer-History.
We quantify this cost using paired differences on the same events in Later TR, whether the new task is solved, and Later ER, whether it is solved while all earlier tests still pass (Section~\ref{sec:stats}).
We separately compare the effort required under the two settings. 
The table also reports $E_1$, where both settings start from identical code, as a reference for run-to-run variation.
Figure~\ref{fig:rq1_cost} shows the paired differences in the two resolution metrics across all ten combinations.

\textbf{Every system solves fewer new tasks on its own history.}
Later TR is lower under Agent-History in all ten combinations, by 0.5 to 9.5 points.
Over all 1,996 paired later events, the difference is $-5.6$ points (95\% CI [$-7.2$, $-4.1$]).
At $E_1$, where both settings start from the same code and TR and ER coincide, the difference is $+1.8$ points ([0.0, $+3.7$]).
Run-to-run variation is therefore small and, if anything, favors Agent-History, so the later losses are not an artifact of running each setting separately.
The cost also holds for Agentless, whose fixed pipeline never explores the repository autonomously.

\textbf{The cost compounds along the episode.}
Averaged over combinations, the TR gap grows from 4.0 points at $E_2$ to 7.2 points at $E_3$, and the $E_3$ gap is the larger one in seven of the ten combinations.
At $E_3$, the Agent-History repository carries two earlier agent changes rather than one.

\textbf{Requiring earlier work to survive more than doubles the cost.}
Later ER is lower under Agent-History in all ten combinations, by 3.1 to 21.0 points (the Later ER bars in Figure~\ref{fig:rq1_cost}).
Over all paired later events, ER falls from 33.6\% to 19.6\% ($\Delta$ER $= -14.0$, [$-16.1$, $-11.9$]), and 346 events are resolved only from developer history against 66 only from agent history.
Few trajectories stay correct for long: at most 24\% of Agent-History episodes remain resolved through all three events, and the highest average number of maintained events is 1.05 out of 3.

\textbf{When both settings succeed, effort barely changes.}
On the 325 events resolved under both settings, the median log token ratio is $+0.03$ ([$-0.03$, $+0.08$]) and the median log ratio of model requests is 0.
OpenHands is the only exception: it uses a median of 19\% more tokens on its own history (log ratio $+0.18$, [$+0.03$, $+0.24$]).
The cost of Agent-History lies in whether an agent succeeds, not in how cheaply it does so.

\begin{tcolorbox}[
  enhanced,
  colback=gray!15,
  colframe=gray!15,
  boxrule=0pt,
  arc=0pt,
  left=6pt,
  right=6pt,
  top=4pt,
  bottom=4pt,
  borderline west={2pt}{0pt}{gray!70}
]
\textbf{Answer to RQ1.}
Evaluating agents from developers' code overstates what they sustain in continuous maintenance.
Building on their own history lowers the success of every coding agent on both backbone models, and the drop grows with each inherited change.
The drop is in how often agents succeed, not in how much work they spend when they do: over the 1{,}996 paired later events, Later TR falls by 5.6 points and Later ER by 14.0 points under Agent-History, yet on the 325 events that both settings resolve, agents use only about 3\% more tokens (median log token ratio $+0.03$). 

\end{tcolorbox}

\subsection{RQ2. Inherited Test Failures or Code Debt?}
\label{sec:rq2}

RQ1 establishes the cost; RQ2 asks where it comes from.
We split the 1,996 paired later events by whether their Agent-History starting code passes all tests of earlier events.
For $E_2$, this means the code after the agent's $E_1$ change passes $T_1$; for $E_3$, the code after its $E_2$ change passes $T_1 \cup T_2$.
Table~\ref{tab:rq2_correctness} reports ER and TR for each cohort.
Because cohort membership depends on how well the agent did earlier, the cohorts differ in difficulty; within each cohort, however, both settings are compared on the same events.

\begin{table}[t]
\centering
\caption{\textbf{Subsequent maintenance by whether the Agent-History starting code passes all earlier tests.}
ER and TR are percentages; $\Delta$ is Agent-History minus Developer-History in percentage points, with 95\% repository-clustered bootstrap intervals.
D-only/A-only count events resolved (ER) under only Developer-History or only Agent-History.}
\label{tab:rq2_correctness}

\small
\setlength{\tabcolsep}{2.6pt}
\renewcommand{\arraystretch}{1.12}

\begin{adjustbox}{max width=\columnwidth}
\begin{tabular}{@{}lrccccc@{}}
\toprule
\textbf{Cohort} & \textbf{$N$} & \textbf{Dev.\ ER} & \textbf{Agent ER} & \textbf{$\Delta$ER [95\% CI]} & \textbf{$\Delta$TR [95\% CI]} & \textbf{D-/A-only} \\
\midrule
All later events            & 1,996 & 33.6 & 19.6 & $-14.0$ [$-16.1$, $-11.9$] & $-5.6$ [$-7.2$, $-4.1$] & 346 / 66 \\
\addlinespace[2pt]
Earlier tests fail          & 1,349 & 23.7 & 1.9  & $-21.9$ [$-24.5$, $-19.3$] & $-7.6$ [$-9.7$, $-5.6$] & 308 / 13 \\
Earlier tests pass          & 647   & 54.3 & 56.6 & $+2.3$ [$-0.9$, $+5.5$]    & $-1.4$ [$-4.0$, $+1.1$] & 38 / 53 \\
\quad + task not pre-solved & 619   & 54.3 & 57.0 & $+2.7$ [$-0.5$, $+6.1$]    & $-1.1$ [$-3.8$, $+1.6$] & 36 / 53 \\
\bottomrule
\end{tabular}
\end{adjustbox}
\vspace{-10px}
\end{table}

\textbf{Two-thirds of Agent-History repositories start with inherited test failures.}
For 1,349 of the 1,996 later events (68\%), the code the agent starts from already fails tests of earlier events.
Classifying the failing test output of the preceding event shows what these failures are.
In 28\% of them, the agent produced no valid change for that event.
In 38\%, earlier tests fail only on assertions, so the inherited code runs but behaves incorrectly.
In 26\%, the failures involve missing names, missing attributes, or mismatched signatures, and the remaining 9\% cannot be classified from the recorded output.

\textbf{Inherited failures are almost never repaired, and they also hinder the new task.}
Only 1.9\% of these events end with all accumulated tests passing under Agent-History, against 23.7\% under Developer-History ($\Delta$ER $= -21.9$).
Part of this gap follows from the definition of ER, which counts the inherited failures against the event.
The current task alone does not: on the same events, agents solve the new task 7.6 points less often from their own history ([$-9.7$, $-5.6$]).
The gap remains when the inherited failures consist of failed assertions only ($-4.4$, [$-7.7$, $-1.0$]), so it does not depend on interface mismatches.
An inherited failure is thus both a debt that the next agent does not pay and an obstacle to its own task.

\textbf{Without inherited failures, the cost disappears.}
On the 647 events whose Agent-History starting code passes all earlier tests, ER is 54.3\% from developer history and 56.6\% from agent history ($\Delta$ER $= +2.3$, [$-0.9$, $+5.5$]), and the new task alone is solved about equally often ($\Delta$TR $= -1.4$, [$-4.0$, $+1.1$]).
One alternative explanation is that earlier agent changes sometimes implement part of a later task in advance.
We test this by running each later event's Fail-to-Pass tests on its Agent-History starting code before the agent runs: only 10 events (1.5\%) already pass, and 18 cannot be evaluated because of environment errors.
Removing both leaves 619 events with $\Delta$ER $= +2.7$ ([$-0.5$, $+6.1$]; McNemar $p = 0.09$), which excludes a disadvantage larger than half a point.
Agent-written code that keeps earlier tests passing is, for the next agent, as good a starting point as the developers' own.

\begin{tcolorbox}[
  enhanced,
  colback=gray!15,
  colframe=gray!15,
  boxrule=0pt,
  arc=0pt,
  left=6pt,
  right=6pt,
  top=4pt,
  bottom=4pt,
  borderline west={2pt}{0pt}{gray!70}
]

\textbf{Answer to RQ2.}
The debt that coding agents leave behind lies mainly in inherited test failures rather than in harder-to-maintain code. 
Agent-written code that keeps earlier tests passing shows no measurable disadvantage as a starting point compared with the developers' own, whereas inherited failures are rarely repaired and also reduce success on the new task.
\end{tcolorbox}

\subsection{RQ3. What Makes an Earlier Change Help or Hinder?}
\label{sec:rq3}

RQ2 shows no aggregate cost once inherited test failures are removed, but individual events still go both ways: in the 619 events of that cohort, Agent-History alone succeeds in 53 and Developer-History alone in 36.
RQ3 asks which properties of the agent's earlier change separate these cases.

We analyze three outcomes on these 619 events.
\emph{Facilitation}: among the 283 events that Developer-History does not resolve, does Agent-History resolve them (53 cases)?
\emph{Hindrance}: among the 336 events that Developer-History resolves, does Agent-History fail (36 cases)?
\emph{Effort}: on the 300 events resolved under both settings, the log ratio of tokens and of model requests.

We measure six properties of the agent's earlier change, each relative to the developers' change for the same event: its excess changed lines, its divergence from the files the developers changed, its alignment with the functions that the next task's developer patch modifies, the cognitive complexity of the changed code, the number of affected dependents, and its alignment with the repository's historical co-change patterns.
Future-Function Alignment uses the next developer patch and is therefore a post hoc characterization, not information available to an agent in advance.
For each outcome and property, we fit a separate mixed-effects model on the standardized property, with repository and target-task random intercepts and controls for approach, model, event position, repository size, earlier change size, and next-task size.
Because the facilitation and hindrance outcomes are sparse, maximum-likelihood fits are unstable, so we fit Bayesian models with weakly informative priors (Normal(0, 1) on coefficients, Student-$t$(3, 0, 1) on group-level standard deviations; 4 chains of 4,000 iterations).
All 24 models satisfy $\hat{R}\leq1.01$ with no divergent transitions.

\begin{table*}[t]
\centering
\caption{\textbf{Associations between properties of the earlier agent change and the next task (619 events whose Agent-History starting code passes all earlier tests).}
Values are Bayesian mixed-effects coefficients for standardized properties with 95\% credible intervals.
For Facilitation and Hindrance, positive values mean higher log-odds of the outcome.
For Token Effort, positive values mean more tokens under Agent-History than under Developer-History.
Bold values have credible intervals that exclude zero.}
\label{tab:rq3}

\small
\setlength{\tabcolsep}{5pt}
\renewcommand{\arraystretch}{1.12}

\begin{tabular}{@{}lccc@{}}
\toprule
\textbf{Property of the earlier change}
& \textbf{Facilitation}
& \textbf{Hindrance}
& \textbf{Token Effort} \\
\midrule

Excess Changed LOC
& $-0.32$ [$-1.04$, $0.30$]
& $0.18$ [$-0.65$, $0.99$]
& $-0.02$ [$-0.11$, $0.07$] \\

Developer File Divergence
& $-0.14$ [$-0.82$, $0.49$]
& $-0.79$ [$-1.74$, $0.07$]
& $0.03$ [$-0.06$, $0.13$] \\

Future-Function Alignment
& $\mathbf{0.77}$ $\mathbf{[0.14, 1.83]}$
& $0.33$ [$-0.26$, $1.04$]
& $\mathbf{-0.15}$ $\mathbf{[-0.28, -0.03]}$ \\

Change-Local Cognitive Complexity
& $\mathbf{-0.94}$ $\mathbf{[-2.51, -0.13]}$
& $-0.35$ [$-2.24$, $1.65$]
& $0.09$ [$-0.91$, $1.08$] \\

Affected Dependents
& $-0.22$ [$-0.85$, $0.32$]
& $0.39$ [$-0.42$, $1.72$]
& $-0.01$ [$-0.08$, $0.05$] \\

Co-change Alignment
& $-0.07$ [$-0.74$, $0.64$]
& $0.60$ [$-0.48$, $1.76$]
& $-0.04$ [$-0.13$, $0.05$] \\

\bottomrule
\end{tabular}
\vspace{-10px}
\end{table*}

\textbf{Earlier changes help when they reach the code the next task needs.}
Of the 24 coefficient intervals, four exclude zero, and three of these belong to Future-Function Alignment.
A one-standard-deviation increase in alignment is associated with 2.2 times higher odds of facilitation (0.77, [0.14, 1.83]), 14\% fewer tokens ($-0.15$, [$-0.28$, $-0.03$]), and 7\% fewer model requests ($-0.07$, [$-0.14$, $-0.003$]).
Because pre-solved tasks are excluded, the association does not come from earlier changes that already complete the next task.
When the agent's earlier change has already touched functions that the next task must modify, the next agent has less to locate and reshape.

\textbf{Simpler changed code helps only with facilitation.}
Lower change-local cognitive complexity relative to the developers' change is associated with facilitation ($-0.94$, [$-2.51$, $-0.13$]), but not with hindrance or effort.

\textbf{No measured property marks agent code as harder to build on.}
For hindrance, all six intervals include zero.
Neither the size of the agent's change, nor its divergence from the developers' files, nor the number of dependents it affects is associated with any of the outcomes.
This is consistent with RQ2: once earlier tests pass, the properties that would signal code debt do not predict when an agent's history hurts the next task.

\begin{tcolorbox}[
  enhanced,
  colback=gray!15,
  colframe=gray!15,
  boxrule=0pt,
  arc=0pt,
  left=6pt,
  right=6pt,
  top=4pt,
  bottom=4pt,
  borderline west={2pt}{0pt}{gray!70}
]

\textbf{Answer to RQ3.}
What makes an earlier change a better foundation is where it lands, not how large or how different it is: changes that reach the code the next task needs make that task both more likely to succeed and cheaper.
None of the measured code properties usually associated with technical debt is linked to worse performance on the next task, further supporting the RQ2 finding that agent-written code passing all earlier tests shows no measurable disadvantage for subsequent maintenance.
\end{tcolorbox}

\section{Discussion}
\label{sec:discussion}

\subsection{Case Studies}

\begin{figure}[t]
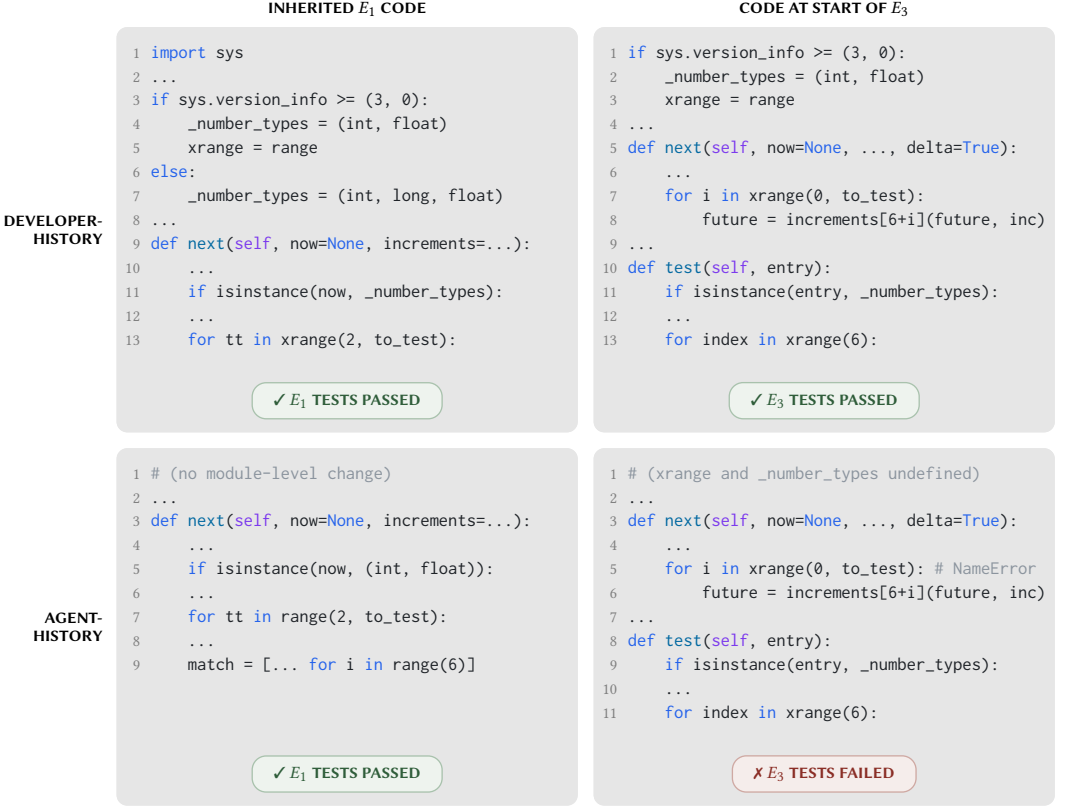

\vspace*{8pt}
\begin{tcbraster}[raster columns=2, raster left skip=1.5cm, raster equal height=rows, raster column skip=6pt, raster row skip=6pt]
\begin{casecode}[colhead={Inherited $E_1$ code}, rowlabel={DEVELOPER-\\HISTORY}]{\testspassed{$E_1$}}
import sys
...
if sys.version_info >= (3, 0):
    _number_types = (int, float)
    xrange = range
else:
    _number_types = (int, long, float)
...
def next(self, now=None, increments=...):
    ...
    if isinstance(now, _number_types):
    ...
    for tt in xrange(2, to_test):
\end{casecode}
\begin{casecode}[colhead={Code at start of $E_3$}]{\testspassed{$E_3$}}
if sys.version_info >= (3, 0):
    _number_types = (int, float)
    xrange = range
...
def next(self, now=None, ..., delta=True):
    ...
    for i in xrange(0, to_test):
        future = increments[6+i](future, inc)
...
def test(self, entry):
    if isinstance(entry, _number_types):
    ...
    for index in xrange(6):
\end{casecode}
\begin{casecode}[rowlabel={AGENT-\\HISTORY}]{\testspassed{$E_1$}}
# (no module-level change)
...
def next(self, now=None, increments=...):
    ...
    if isinstance(now, (int, float)):
    ...
    for tt in range(2, to_test):
    ...
    match = [... for i in range(6)]
\end{casecode}
\begin{casecode}{\testsfailed{$E_3$}}
# (xrange and _number_types undefined)
...
def next(self, now=None, ..., delta=True):
    ...
    for i in xrange(0, to_test): # NameError
        future = increments[6+i](future, inc)
...
def test(self, entry):
    if isinstance(entry, _number_types):
    ...
    for index in xrange(6):
\end{casecode}
\end{tcbraster}
\caption{\texttt{parse\_crontab} (Claude Code with DeepSeek-V4-Flash). Two changes that are both functionally correct need not lead to equal maintenance outcomes. An example instance where both the developer's and the agent's change pass the tests of the initial task and the next one. After later upstream changes, the version built on the developer's code passes, while the version built on the agent's code fails on a task unrelated to the original change.}
\label{fig:case_crontab}
\vspace{-10pt}
\end{figure}

\begin{figure}[t]
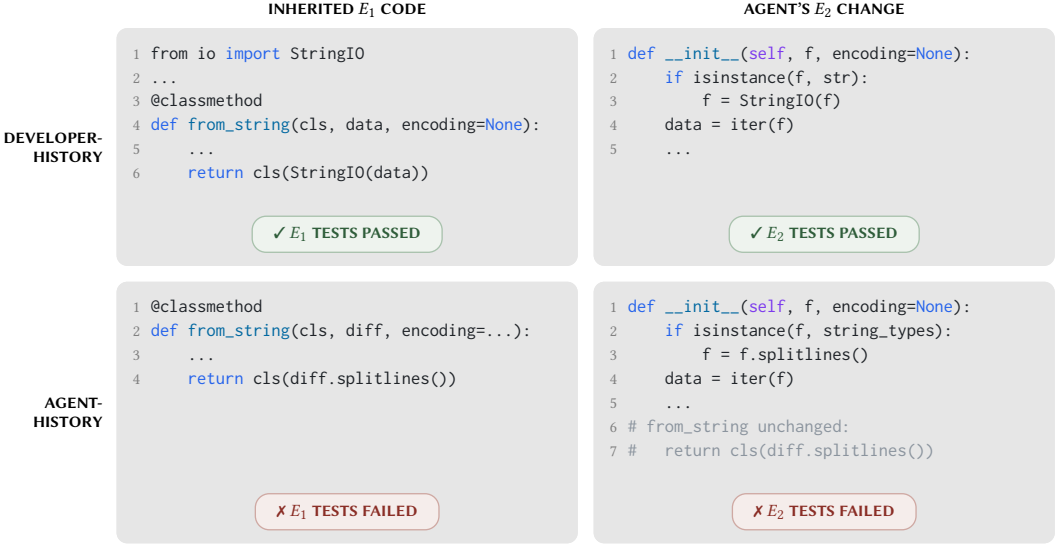

\vspace*{8pt}
\begin{tcbraster}[raster columns=2, raster left skip=1.5cm, raster equal height=rows, raster column skip=6pt, raster row skip=6pt]
\begin{casecode}[colhead={Inherited $E_1$ code}, rowlabel={DEVELOPER-\\HISTORY}]{\testspassed{$E_1$}}
from io import StringIO
...
@classmethod
def from_string(cls, data, encoding=None):
    ...
    return cls(StringIO(data))
\end{casecode}
\begin{casecode}[colhead={Agent's $E_2$ change}]{\testspassed{$E_2$}}
def __init__(self, f, encoding=None):
    if isinstance(f, str):
        f = StringIO(f)
    data = iter(f)
    ...
\end{casecode}
\begin{casecode}[rowlabel={AGENT-\\HISTORY}]{\testsfailed{$E_1$}}
@classmethod
def from_string(cls, diff, encoding=...):
    ...
    return cls(diff.splitlines())
\end{casecode}
\begin{casecode}{\testsfailed{$E_2$}}
def __init__(self, f, encoding=None):
    if isinstance(f, string_types):
        f = f.splitlines()
    data = iter(f)
    ...
# from_string unchanged:
#   return cls(diff.splitlines())
\end{casecode}
\end{tcbraster}
\caption{\texttt{python\_unidiff} (Claude Code with DeepSeek-V4-Flash). Inherited test failures are rarely repaired and also hinder the next task. An example instance where the agent's first change fails its own tests. When the same agent makes a subsequent change, the version built on the developer's code passes, while the version built on its own code extends the earlier defect instead of repairing it, and fails.}
\label{fig:case_unidiff}
\vspace{-12pt}
\end{figure}

We include two instances in which an agent solves a later event when it builds on the developers' earlier patches (Developer-History) but fails the same event when it builds on its own earlier changes (Agent-History) (Figs.~\ref{fig:case_crontab} and~\ref{fig:case_unidiff}).

In \texttt{parse\_crontab} (Claude Code with DeepSeek-V4-Flash; Fig.~\ref{fig:case_crontab}), the agent is first asked to make \texttt{CronTab} work under Python~3. The developer adds a module-level alias that rebinds \texttt{xrange} and defines the accepted number types once for the whole module; the agent instead rewrites only the Python~2 call sites inside \texttt{next()} that the task exercises. Both changes pass the tests of this event and of the next one. The project, however, keeps evolving: later commits from the repository's history add new code that relies on the module-wide alias. When the agent is later asked to accept \texttt{-sun} as the end of a weekday range, the alias covers this code under Developer-History, and the agent solves the task. Under Agent-History, these names are undefined, and 8 of 10 tests fail with a \texttt{NameError} in code unrelated to the task. The agent's change thus passes the current tests but fixes only the symptoms they exercise, not the module-wide requirement that later maintenance relies on.

In \texttt{python\_unidiff} (Claude Code with DeepSeek-V4-Flash; Fig.~\ref{fig:case_unidiff}), the agent is first asked to add a \texttt{PatchSet.from\_string} helper.
Under Developer-History, the developer wraps the string in \texttt{StringIO} so that it is parsed like an open file; the agent instead splits it into lines, which drops the line terminators, and the hidden tests of this event fail. Because the agent never sees these tests, the failure is carried forward. In its next event, the agent is asked to let the \texttt{PatchSet} constructor accept strings. Under Developer-History, it converts strings with \texttt{StringIO} and passes. Working from its own earlier change, the agent reads the line-splitting logic it wrote and reuses it in the constructor, so the inherited failure stays unrepaired and the new task fails on the same mismatch.

Both cases share the failure mode that RQ2 measures at scale: a change that satisfies the tests in front of it leaves a requirement unmet, and the next agent builds on the code its predecessor left rather than on the tests that code fails.

\subsection{Implications}

\subsubsection{For evaluating and measuring continuous maintenance}

Standard evaluation runs each task from the developers' code and judges it only by its own tests.
Our results show what this misses: the loss in Event Resolve under Agent-History is two and a half times the loss in task resolution (RQ1), so task resolution alone scores a system on the task in front of it, not on the repository it hands back.
Evaluations should therefore report outcomes that accumulate across a sequence, such as Event Resolve and Average Maintained Events, and run each later task from the agent's own preceding changes alongside the developers' code; the direction of the difference held in all ten combinations we tested.
Static properties of the returned code are no substitute: none of the structural properties in RQ3 predicts when an earlier change hinders the next task, whereas whether earlier tests still pass does.
Future benchmarks should also measure failure propagation directly, that is, how long an unrepaired failure survives and whether later agents repair it, which requires chains longer than the three events released here. 

\subsubsection{For designing coding agents}

The cost lies in inherited test failures rather than in agent-written code itself, and it persists even when the inherited code runs but behaves incorrectly (RQ2).
Three design directions follow.
First, agents should run the tests of earlier tasks before declaring a task done; in our harness these tests are hidden, so an agent cannot see that it broke earlier behavior, whereas an agent that runs them can repair a failure at the event that caused it, while the repair is still local.
Second, passing the current tests is necessary but not sufficient: in the crontab case (Fig.~\ref{fig:case_crontab}), a change that fixed only the tested call sites passed two consecutive events and still broke code added later.
Third, an agent that inherits a failure should repair it rather than build on it; in the unidiff case (Fig.~\ref{fig:case_unidiff}), the agent instead reused its own defective logic and failed the new task on the same mismatch. 

\subsubsection{For practitioners}

Practitioners adopt coding agents into repositories that outlive any single change, so what matters is the state an agent leaves behind, not the change it reports: at most 24\% of Agent-History episodes stay fully resolved through all three events (RQ1).
The practical safeguard is to keep the tests of earlier tasks in the pipeline and run them after every agent change, not only the tests the agent was asked to satisfy.
Conversely, when earlier tests still pass, agent-written code is as good a starting point as the developers' (RQ2), so protecting later maintenance requires keeping those tests passing, not reviewing agent code for style or structural drift. 

\section{Threats to Validity}
\label{sec:threats}

\textbf{Internal validity.}
Each setting is run once per event, so individual paired outcomes vary by chance.
We draw conclusions only from aggregate paired differences with repository-clustered intervals, and the $E_1$ comparison, where both settings start from identical code, bounds this variation at $+1.8$ points ([0.0, $+3.7$]), well below the later-event costs.
Developer-History starts from historical code that may appear in model pretraining data, which could favor it.
Such an advantage would also appear on events whose Agent-History code passes all earlier tests, where we observe no disadvantage (Section~\ref{sec:rq2}).
The RQ2 cohorts are defined by how well the agent did on earlier events and therefore differ in difficulty; we compare the two settings only within a cohort, on the same events, and report the full population alongside.
Future developer patches and tests are used only after evaluation, to detect pre-solved tasks and compute RQ3 properties, and are never shown to agents.

\textbf{Construct validity.}
Hidden tests are not complete specifications, so inherited test failures and their absence refer only to tested behavior.
Because the tests come from the developers' history, some encode the developers' interfaces, and 26\% of inherited failures involve missing names, attributes, or mismatched signatures.
The cost to the new task remains when inherited failures are assertion failures only ($\Delta$TR $= -4.4$, [$-7.7$, $-1.0$]), so it does not depend on interface mismatches.
Instructions for 137 of the 300 events are restated from short commit titles and may underspecify the task.
Both main results hold separately for these events and for the others: the cost over all later events is $-12.3$ and $-15.5$ points in $\Delta$ER, and the cohort that passes all earlier tests shows $+1.8$ ([$-3.1$, $+6.9$]) and $+3.3$ ([$-0.5$, $+7.4$]).
Episodes replay selected developer patches without intermediate commits, and 108 events use tests written from the instruction; both settings share the same replayed history, and all tests pass the same certification (Section~\ref{sec:certification}).
Our conclusion about code debt rests on paired maintenance outcomes, not on static maintainability metrics, whose validity as proxies is debated.

\textbf{External validity.}
SWE-Maintain covers 100 three-event episodes from 50 Python repositories, five coding systems, and two backbone models.
Other languages and development practices may differ, and longer sequences may compound the cost further, as its growth from $E_2$ to $E_3$ suggests.
The results are consistent across all ten system--model combinations, including the non-agentic Agentless pipeline.
In Agent-History, the same system continues from its own changes; we do not study agents that build on another agent's work or on mixed human--agent histories.

\textbf{Conclusion validity.}
RQ3 relates six properties to three outcomes, and its facilitation and hindrance outcomes rest on 53 and 36 events.
We therefore treat RQ3 as exploratory, report results for all six
properties rather than selecting properties post hoc, and interpret
only the property whose association recurs across outcomes.
These associations are observational; Future-Function Alignment in particular uses the next developer patch and describes, rather than predicts, which earlier changes help.

\section{Related Work}

\textbf{Evaluating coding agents.}
SWE-bench evaluates whether agents resolve real GitHub issues against executable tests~\cite{jimenez2024swe}, SWE-bench Verified and Multi-SWE-bench refine its task quality and extend it to other languages~\cite{openai2024swebenchverified,zan2025multiswebench}, and other benchmarks target feature development and its reasoning~\cite{zhou2026featurebench,liu2026racebench}, the full issue-resolution cycle~\cite{guan2026swecycle}, and the generation of whole applications or repositories~\cite{pan2026ralbench,ding2026nl2repobench}.
Each of their tasks starts from code that developers wrote, which is our Developer-History setting, and our results show that this setting overstates what agents sustain once they work on their own code.
Studies of SWE-bench have also found that some patches judged resolved pass weak tests while remaining incorrect, and that some tasks leak their solution~\cite{aleithan2024swebenchplus,wang2025solved}; SWE-Maintain addresses the corresponding risk for sequences by certifying that the developers' own history passes every accumulated test after every event.
Closest to our setting, a growing set of benchmarks evaluates agents over sequences of related tasks: SWE-Bench-CL orders issues chronologically to study continual learning~\cite{joshi2025swebenchcl}, SWE-EVO derives long-horizon evolution tasks from real histories~\cite{thai2025sweevo}, SWE-CI studies sustained modification under continuous integration~\cite{chen2026sweci}, SWE-Milestone builds later milestones on earlier work~\cite{deng2026swe}, SWE-Chain chains package upgrades so that each step inherits the previous codebase~\cite{lam2026swechain}, and ChainSWE poses dependent bug fixes in the same repository~\cite{jin2026chainswe}.
These benchmarks establish that agents must be evaluated over evolving code, but in a sequence a later task's outcome reflects both the task's difficulty and the code it starts from.
SWE-Maintain separates the two by running each later task from both the developers' and the agent's history, and it complements these benchmarks: any of them can add developer-history runs of its own tasks to obtain the same paired estimate.

\textbf{Technical debt, code quality, and regression.}
Technical debt names the future cost of changes that satisfy an immediate need~\cite{cunningham1992wycash,kruchten2012technical}; it has been classified into types such as code, design, and test debt~\cite{alves2016identification}, identified through developers' own comments~\cite{potdar2014exploratory}, integrated with quality management~\cite{ramasubbu2018integrating}, and studied in the co-evolution of production and test code~\cite{guo2025understanding}.
Software evolution research holds that systems become harder to change as changes accumulate~\cite{lehman1980programs} and relates later maintenance burden to architectural complexity, change dynamics, and refactoring~\cite{cai2025understanding,geipel2009software,noei2025empirical}, and studies of AI-generated code have measured its correctness, maintainability issues, and code smells~\cite{yetistiren2022assessing,liu2024refining} as well as pull requests authored by autonomous coding agents~\cite{li2025aiteammates}.
Program repair research has shown that patches can pass tests while remaining incorrect~\cite{qi2015analysis,smith2015cure} and has developed techniques to detect such overfitting and strengthen validation tests~\cite{xiong2018identifying,xin2017identifying,yang2017better}, while regression testing selects and prioritizes tests that catch a change breaking earlier behavior~\cite{rothermel1997safe,yoo2012regression}.
These lines judge code through indicators of the code itself, or judge a single change at the time it is made.
We instead measure the debt of agent-written code by its consequence, the paired change in what later agents achieve on it, after the change has been accepted.
This separates two sources that code-level indicators cannot: code debt is not detectable in agent code that passes earlier tests, and none of the static properties we examine predicts when such code hinders the next task, whereas inherited test failures carry most of the cost, are almost never repaired, and compound along an episode, which makes regression checks across an agent's successive changes as important as the correctness of each change.

\section{Conclusion}

This paper studies whether building on its own earlier changes makes a coding agent's subsequent maintenance harder, and if so, where the cost comes from.
We make three contributions.
First, we propose a paired evaluation design that runs each later maintenance task twice, once from the developers' history, as standard benchmarks do, and once from the agent's own, so that any difference reflects the inherited code rather than the task.
Second, we build SWE-Maintain, 100 behaviorally certified three-event episodes from 50 real-world Python repositories, which instantiates this design.
Third, our study of five coding systems and two backbone models shows that the cost is real and comes from inherited test failures rather than from the code the agent writes.
Our findings have several implications. 
For example, evaluations that start every task from developers' code overstate how well agents sustain maintenance, so benchmarks should also run agents from their own history.
Agents, in turn, should treat keeping earlier tests passing as part of every change, because the debt they accumulate is unrepaired test failures, not code that is harder to maintain.

\bibliographystyle{ACM-Reference-Format}
\bibliography{references}

\end{document}